\documentclass[12pt,showpacs,prd,preprint,nofootinbib,preprintnumbers]{revtex4}%

\usepackage{amsmath,amssymb,epsf,epsfig,amsthm,bm,graphicx}
\usepackage[dvips]{color}

\theoremstyle{theorem}
\newtheorem{theorem}{Theorem}[section]

\newtheorem{proposition}[theorem]{Proposition}

\theoremstyle{definition}

\newtheorem{definition}[theorem]{Definition}

\begin{document}
\title{Lovelock gravity and Weyl's tube formula}
\author{Steven Willison$^{1}$} \affiliation{$^{1}$Centro de Estudios
Cient\'{\i}ficos, Casilla 1469, Valdivia, Chile.}
\email{steve-at-cecs.cl} 

\begin{abstract}
In four space-time dimensions, there are good theoretical reasons
for believing that General Relativity is the correct geometrical
theory of gravity, at least at the classical level. If one admits
the possibility of extra space-time dimensions, what would we expect
classical gravity to be like? It is often stated that the most
natural generalisation is Lovelock's theory, which shares many
physical properties with GR. But there are also key differences and
problems. A potentially serious problem is the breakdown of
determinism, which can occur when the matrix of coefficients of
second time derivatives of the metric degenerates. This can be
avoided by imposing inequalities on the curvature. Here it is argued
that such inequalities occur naturally if the Lovelock action is
obtained from Weyl's formulae for the volume and surface area of a
tube. Part of the purpose of this article is to give a treatment of
the Weyl tube formula in terminology familiar to relativists and to
give an appropriate (straightforward) generalisation to a tube
embedded in Minkowski space.
\end{abstract}

{\small CECS-PHY-09/04}

\maketitle

\section{Introduction}
\subsection{Generalising GR to higher dimensions}

The Einstein equation can be cast as a hyperbolic system of
equations. Therefore deterministic evolution of the metric from an
initial geometric data (satisfying the initial value constraints) is
guaranteed provided that spacetime is globally hyperbolic. This is
an important result which establishes GR as a legitimate classical
theory- one might say that determinism is the defining
characteristic of a classical theory of physics.

In dimensions greater than four, there are other symmetric tensors
$H_{\mu\nu}(g_{\mu\nu},$ $ g_{\mu\nu,\rho},$
$g_{\mu\nu,\rho\sigma})$, known as Lovelock
tensors\cite{Lovelock-71}, that one can add which satisfy an
identity $\nabla_\mu H^\mu_\nu =0$ derived from the Bianchi
identity. Therefore, if we add these tensors to the Einstein
equation, we still expect to have just the right number of
independent equations to determine the metric up to diffeomorphisms.
Furthermore the Lovelock tensors are second order in derivatives, so
one expects to have the same initial data (which we may assume to be
the spatial metric and its first time derivative). The Lovelock
tensors are polynomials in the curvature of the form:
\begin{gather}
 (H^{(n)})^\mu_\nu := \frac{-1}{2^{n+1}} \delta^{\mu \rho_1 \cdots \rho_{2n}}
 _{\nu \sigma_1 \cdots \sigma_{2n}} R^{\sigma_1\sigma_2}_{\quad
 \rho_1 \rho_2} \cdots R^{\sigma_{2n-1}
 \sigma_{2n}}_{\quad \rho_{2n-1}\rho_{2n}}\, .
\end{gather}
The equation of Lovelock gravity in $D = 2m+1$ or $2m+2$ dimensions
will be
\begin{equation}\label{Lovelock_field_eqn}
 \Lambda g_{\mu\nu} + G_{\mu\nu} + \alpha_2 (H^{(2)})_{\mu\nu} +
 \cdots \alpha_m (H^{(m)})_{\mu\nu} = \kappa T_{\mu\nu}
\end{equation}
with $(H^{(m)})_{\mu\nu}$ being the highest order term which does
not vanish identically.

Lovelock gravity has been studied in various contexts:
compactified\cite{Deruelle-89} and brane-world\cite{Charmousis-02}
cosmological models
%(some more recent works are
%\cite{Dufaux-04}\cite{Cai-05})
; Kaluza-Klein theory\cite{MuellerHoissen-85}\cite{Ishikawa-86} (a
more recent work is \cite{Canfora-08}); black
holes\cite{Boulware-85}\cite{Wheeler-85}; Chern-Simons theories of
(super)-gravity\cite{Chamseddine-89}\cite{Chamseddine-90}\cite{Troncoso:1998ng},
to name but some. Mathematical properties of the Lovelock terms have
been studied in Refs. \cite{Patterson-81} and \cite{Labbi-07}.

\subsection{Determinism and curvature inequalities}

However, there is a problem which afflicts Lovelock's theory. This
problem was encountered some time ago by Teitelboim and
Zanelli\cite{Teitelboim-87} working in the Hamiltonian formalism and
by Choquet-Bruhat who considered the Cauchy
problem\cite{Choquet-Bruhat-88,Choquet-Bruhat-88b}. Here we shall
briefly review the problem, following Ref.
\cite{Choquet-Bruhat-88b}. Let us introduce a time-slicing, writing
the metric in ADM form and setting the shift to zero. Let $h_{ab}$
be the intrinsic metric on the constant-time hypersurface. The
curvature component containing the second time derivatives is $R_{\
a0b}^0 \simeq \frac{1}{2\alpha^2} g_{ab,00}$ where $\alpha$ is the
lapse function and $\simeq$ means equality modulo terms of lower
order in time derivatives. Looking at the field equations (in empty
space), one finds that $H^0_0 =0$ and $H^0_a =0$ contain only first
time derivatives and therefore will be initial-value constraints;
$H^a_b =0$ contains terms $g_{ab, 00}$ and therefore describes the
evolution of the system. The relevant part of the Lovelock tensors
is
\begin{gather*}
 (H^{(n)})_{ab} \simeq \frac{1}{\alpha^2}(\Xi^{(n)})_{ab}^{\ \ cd}\, g_{cd,00}\, ,
 \qquad
 (\Xi^{(n)})_{ab}^{\ \ cd} := \frac{-n}{2^{n}}
  g^{fd} g_{ae} \delta^{e c b_3 \cdots b_{2n}}
 _{b f a_3 \cdots a_{2n}} R^{a_3 a_4}_{\quad
 b_3 b_4} \cdots R^{a_{2n-1}a_{2n}}_{\quad b_{2n-1} b_{2n}}\, .
\end{gather*}
It is helpful to use the trace of the equations to cast them in the
form $R_{ab} +$ Lovelock corrections. Then we get
\begin{gather}
 \frac{1}{2\alpha^2} \left( \delta_a^c \delta_b^d -
  \mathcal{Y}_{ab}^{\ \ cd}\right) g_{cd,00} =
  f_{ab}(g,\dot{g}, g' , g'' ,... ) \simeq 0\, ,
 \\
 \mathcal{Y}_{ab}^{\ \ cd}:=
 \sum_{n=2}^{[d-1/2]} 2 \alpha_n \left( (\Xi^{(n)})_{ab}^{\ \ cd} -
 \frac{1}{D-2}g_{ab} g^{ef} (\Xi^{(n)})_{ef}^{\ \ cd}\right)\, .
\end{gather}
(Note that $\alpha$ is not a dynamical variable, which corresponds
to the fact that locally one can always choose Gaussian normal
coordinates). Above, the term $f_{ab}$ denotes a matrix which
depends only on the initial data $g_{ab}$, $g_{ab,0}$ and spatial
derivatives, but not on $g_{ab,00}$.

It is useful to combine the symmetrised pairs of indices into a
single index $I := ab$, $J: = cd$, so that $\delta_I^{\ J} +
\mathcal {Y}_I^{\ J}$ is a $d(d+1)/2$-by-$d(d+1)/2$ matrix. The
system is solvable for $g_{cd,00}$ iff \begin{equation*}\det
(\delta_I^{\ J} + \mathcal{ Y}_I^{\ J}) \neq 0\, .\end{equation*}
Unlike for Einstein's theory, in Lovelock gravity those coefficients
are functions, and so it may be that the determinant is non-zero in
some regions but vanishing in other regions. At such points where
the determinant vanishes, there is an ambiguity of the continuation
of space-time into the future.\footnote{ In considering the Cauchy
problem in this way, one treats the intrinsic metric and its time
derivative (i.e. extrinsic curvature) as the initial data. In the
Hamiltonian approach one has $g_{ab}$ and the canonical momenta
$\Pi_{ab}$. Hamilton's equation is of the form $\dot \Pi_{ab} =
...$. However, non-determinism enters when one faces the fact that
one can not always invert $\dot{g}_{ab}$, appearing on the r.h.s.,
to express it as a function of $\Pi_{ab}$. Of course
$\det(\delta_I^{\ J} + \mathcal{ Y}_I^{\ J}) =0 \Leftrightarrow \det
\left( \frac{\partial \Pi_I}{\partial K_J}\right) =0$, so the
ill-posed Cauchy problem and the breakdown of the Hamiltonian method
are closely related. However they are not quite equivalent. For
example non-invertibility can even occur on a hypersurface in
Minkowski space where $\dot{g}_{ab}$ can jump dramatically without
discontinuity in $\Pi_{ab}$ (this solution was found explicitly in
Ref. \cite{Garraffo-07}). In that case $\det (\delta_I^{\ J} +
\mathcal{ Y}_I^{\ J})$ is certainly not zero. Similar issues are
discussed in Ref. \cite{Deruelle-Madore-03}.
\\
This appears to be related to the results of Ref.
\cite{Miskovic:2007mg} where it was shown that the Hamiltonian
evolution normal to a boundary (in that case at infinity) is
equivalent to the Lagrangian treatment only if additional Dirichlet
boundary terms are added to the Lagrangian.}

If the matrix $\mathcal{Y}_I^{\ J}$ is small, then the determinant
is positive definite and deterministic evolution is guaranteed.
Roughly speaking, this will be true if the curvature components are
small compared with lengthscales$^{-2}$ constructed from the
coupling constants. Is there some interpretation of the theory in
which such inequalities on the curvature  arise naturally?

\section{Weyl's Tube formula}

In what follows, we develope some new ideas concerning the relation
between Weyl's classic formulae for the volume and area of a tube on
the one hand, and Lovelock gravity and the problem of determinism on
the other.

\subsection{Euclidean tube formula}

Let $M$ be a $D$-dimensional submanifold of $R^N$. It's $l$-tube is
defined to be the set of all points in $R^n$ with distance $\leq l$
from $M$ along a geodesic which intersects $M$ normally (if $M$ has
no boundary, this is the same as the set of all points of shortest
distance $\leq l$ from $M$). If $l$ is small enough compared to the
curvature radii of $M$ at every point, then the tube is
diffeomorphic to $M \times B_{N-D}$, where $B_{N-D}$ is the unit
ball of dimension $N-D$. For small enough $l$, a formula due to Weyl
says that the volume of the $l$-tube is:
\begin{equation}\label{Weyl_Tube_Formula}
 V = \text{Vol}(B_{N-D}) \sum_{n=0}^{[D/2]}
     \frac{(N-D)!!}{(N-D +2n)!! (2n)!!} \ { l^{N-D+2n}}
     \int_M {\cal L}^{(n)}\, ,
\end{equation}
where
\begin{equation}
 {\cal L}^{(n)} := \frac{1}{2^{n}} \delta^{ \rho_1 \cdots \rho_{2n}}
 _{\sigma_1 \cdots \sigma_{2n}} R^{\sigma_1\sigma_2}_{\quad
 \rho_1 \rho_2} \cdots R^{\sigma_{2n-1}
 \sigma_{2n}}_{\quad \rho_{2n-1}\rho_{2n}} \sqrt{g} d^Dx\, .
\end{equation}
(See \cite{Gray-04} for an interesting review.) If any of the
curvature radii are small compared to $l$, we expect the formula to
break down because different sections of the tube associated with
different regions of $M$ can intersect. This would cause the formula
to overcount the volume.

It was recently pointed out by Labbi\cite{Labbi-07} that the
curvature invariants appearing in Weyl's formula are the same as
those appearing in the Lagrangian of Lovelock's theory. So the
volume of a tube coincides with the action of Euclidean Lovelock
theory with a special choice of coupling constants. It would be
interesting to generalise the tube formula to a Minkowski space
background. Also, it may be of interest to find a tube formula in
(A)dS space. The generalisation to hyperbolic space is well
known\cite{Gray-04}.

\subsection{Minkowski space tube formula}

The first question which arises in generalising to Minkowski space
is how to define the tube. In the Euclidean case the definition is
motivated by the intuitive fact that the shortest route from a point
to a surface is the line that hits the surface normally. In
Minkowski space this is no longer true. Indeed it would be futile to
define the tube as the locus of points of less than $l$ spacelike
proper distance from $M$ for a simple reason. Let $p$ be a point on
$M$. Then any points which are infinitesimally close to the
lightcone of $p$ and which have spacelike separation from $p$ must
be included in the tube. So a tube thus defined would stretch all
the way out to future and past null infinity. However, even though
the meaning is not quite the same as as in the Euclidean case, we
can still define the tube in the same way:
\begin{definition}\label{Minkowski_Tube_Def}
The tube of $M^d$ in Minkowski space $\mathbb{M}^n$ is the set of
all points of proper distance less than $l$ along a geodesic which
intersects $M^d$ normally.
\end{definition}
According to the above definition, the tube will not extend out
towards null infinity unless the normal vector of $M^d$ becomes null
at some point. So for an embedded submanifold of strictly Minkowski
signature, the tube is bounded.

It is curious that, although the geometry of Minkowskian tubes is
quite different compared to their Euclidean counterparts, the
formula for the volume turns out to be the same. Before considering
the general proof of this, let us check it explicitly with a pair of
examples.

First we consider the embedding of an $(N-1)$-sphere $S_r^{N-1}$
into $\mathbb{R^N}$ and then the Lorentzian equivalent,
 de Sitter space embedded as a hyperboloid in Minkowski space. In
the first case, the volume of the tube is the volume contained
between two concentric spheres of radius $r - l$ and $r+l$, i.e.
\[
V = N \text{Vol}(S^{N-1}) \left\{ (r+l)^N - (r-l)^N\right\}
\]
Using $\int {\cal L}^n = r^{-N + 2n +1}
\frac{(N-1)!}{(N-2n-1)!}\text{Vol}(S^{N-1})$ for the Lovelock
scalars of the sphere of radius $r$ we can expand the tube volume
as:
\begin{equation}\label{sphereTube}
V = 2 \sum_{n=0}^{[(N-1)/2]} \frac{l^{2n+1}}{(2n +1)!}\int_{M} {\cal
L}^{(n)}
\end{equation}
with $M =S_r^{N-1}$.

\begin{figure}
  \includegraphics[width=.35\textwidth]{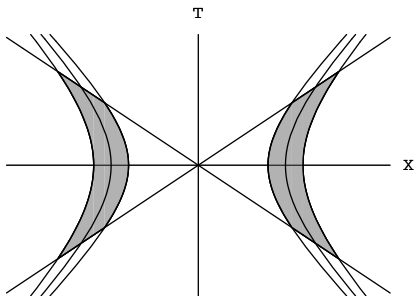}\\
  \caption{{\small The tube of a section of de Sitter space $dS_r^{N-1}$
  embedded in Minkowski
  space $\mathbb{M }^N$. A slice through the $(X,T)$ plane is shown.
  The tube is the shaded region between concentric $dS$ spaces of
  curvature radii
  $r-l$ and $r+l$. The normal vectors are aligned with rays through
  the origin.}}\label{dSFig}
\end{figure}

In the second case, we have $dS^{N-1}_\rho$ realised by the
embedding $-dT^2 + d\vec{X} \cdot d\vec{X} = \rho^2$ in
$\mathbb{M}^N$. It is useful to parametrise this by $T = \rho \sinh
\chi$ etc. As with the sphere, the normal vectors lie along rays
through the origin and one finds that the tube is delimited by two
concentric embedded $dS$ spaces of curvature radii $\rho = r-l$ and
$\rho =r+l$ respectively. The volume element is $\rho^{N-1}d\rho\,
d\Omega_{dS}$ where $d\Omega_{dS}$ is the volume element on the
hyperboloid of unit curvature \footnote{The volume diverges, but we
can restrict to the region $\chi_i \leq \chi \leq \chi_f$. This
correctly accounts for the edge effects of the tube, because the
lines $\chi =$ const. coincide with the normal vectors (see fig
\ref{dSFig}).}.
So we obtain for the volume:
\[
 V = N \text{Vol}(dS^{N-1}) \left\{ (r+l)^N - (r-l)^N\right\}
\]
Since the Lovelock curvature scalars are the same for $dS$ as for
the sphere, we obtain the same formula (\ref{sphereTube}).

Let us now consider the tube of a general Lorentzian manifold $M^D
\subset \mathbb{M}^N$. Since the definition of a tube in terms of
the normal vectors is the same as for the Euclidean case, one would
expect Weyl's formula in terms of extrinsic curvatures\cite{Gray-04}
to be the same. Also, since the Gauss-Coddazzi equations are the
same, we expect the formula (\ref{Weyl_Tube_Formula}) in terms of
intrinsic curvatures to apply also to Minkowski space. In order to
confirm this, let us briefly revisit the proof of the tube formula,
formulating things in a terminology familiar to relativists using
Minkowskian signature.

Let us consider an infinitesimal region on $M^D$ and let $(e_{(1)},
\dots, e_{(D)})$ be a set of orthonormal vectors forming a basis of
the tangent space. In a local neighbourhood, this can be extended to
an orthonormal basis of $T\mathbb{M}^N$ which can also be
interpreted as a set of direction vectors in $\mathbb{M}^N$, denoted
$(\vec{e}_{(1)}, \dots, \vec{e}_{(D)}, \vec{n}_{(D+1)}, \dots,
\vec{n}_{(N)})$, where the $n_{(i)}$ are normal vectors to $M^d$.

Consider an infinitesimal $D$-cube defined by the vectors
$\vec{v}_{(1)}\equiv\theta^{(1)} \vec{e}_{(1)},$ $\dots$, $
\vec{v}_{(D)} \equiv\theta^{(D)} \vec{e}_{(D)}$ where $\theta^{(i)}
= \theta^{(i)}_\mu \delta x^\nu$ are the infinitesimal line elements
in the direction of the vector. Now we displace the vertices of the
cube by a vector $\sum_i z^i \vec{n}_i$ in a normal direction. If
there is extrinsic curvature then $\vec{n}_i$ will vary from one
vertex to another. Therefore the displaced infinitesimal vectors
will be
\begin{gather}
 \vec{v}_a(z) = \vec{v}_a + z^i \nabla_{v_{a}} \vec{n}_i
% = \theta^a + \theta^b x^i e_b^\mu \nabla^{a} N^i_\mu
 = \vec{v}_a - \vec{v}_b  z^i K^b_{a\, i}
\end{gather}
 where $K^{b}_{a\, i}$ is the extrinsic curvature
tensor w.r.t. the normal $\vec{n}_i$. The displaced $D$-volume
element is therefore
%(interpreting $\theta^{a} = \theta^a_\mu dx^\mu$ as dual
%basis vectors)
\begin{equation}
    \det\left(\delta^a_b - z^i K^{a}_{b\, i}\right)
  \theta^{(1)} \wedge \cdots \wedge \theta^{(d)}\, .
\end{equation}
Integrating these elements over $z^i$ and over $M^D$, assuming that
they do not intersect each other, gives:
\begin{equation}
  \int_{M^D} \int_{z^i z^i < l^2} \det\left(\delta^a_b - z^i K^{a}_{b\, i}\right)
   dz^{D+1} \cdots dz^{N} dV(M^D)
  \, .
\end{equation}
The calculation of combinatorial factors amounts to calculating the
moments of the $(N-D)$-ball, $\langle z^i\rangle$, $\langle z^i
z^j\rangle$ etc. Since odd moments vanish, the extrinsic curvatures
will always appear in pairs, $(K^a_{c\, i} K^{b}_{d\, i} - K^b_{c\,
i} K^{a}_{d\, i}) \cdots$ when we expand out the determinant. In
this way, the extrinsic curvatures can always be substituted for
intrinsic curvatures using the Gauss formula. Since the normal space
is Euclidean, the Gauss formula is the same as in the Euclidean case
$R^{ab}_{\ \ cd} = \sum_i (K^a_{c\, i} K^{b}_{d\, i} - K^b_{c\, i}
K^{a}_{d\, i})$. Therefore, the same combinatorial factors and signs
must arise\footnote{For details of the calculation of the
combinatorial coefficients, see \cite{Gray-04}. Alternatively,
following Weyl, one can take an example where $M^D$ is of constant
curvature, and read off the coefficients.}. So we conclude:

\begin{proposition}
Let $M^D$ be a manifold with metric of Minkowski signature embedded
in Minkowski space $\mathbb{M}^N$. Let $T(M)$ be the $l$-tube of
$M^D$. Assuming that every point in the tube has a unique geodesic
which connects it with $M^D$ and intersects $M^D$ normally, the
world-volume of the tube is correctly described by formula
(\ref{Weyl_Tube_Formula}).
\end{proposition}

\subsection{Example in D=5 illustrating the failings of the volume
formula}

In this case we have:
\begin{gather}
 V = \frac{\text{Vol}(B_{N-5})\, l^{N-3}}{ 2 (N-3) }
 \int_{M^5} \!\left( -2\Lambda + R + \alpha(R^2-4R_{\mu\nu}R^{\mu\nu}
 + R_{\mu\nu\rho\sigma}R^{\mu\nu\rho\sigma})\right)\sqrt{-g} d^5x
\end{gather}
with
\begin{gather}
 \Lambda = -\frac{N-3}{l^2} \, , \qquad
 \alpha = \frac{l^2}{4 (N-1)}\, .
\end{gather}
We note that the ``magic" combination of coupling constants has the
value $
 x:= \frac{4 \alpha \Lambda}{3} = - \frac{N-3}{3(N-1)}$.
For all dimensions $N > 5$ this is in the range $-1/3 < x \leq
-1/5$. The Chern-Simons gravity theories correspond to $x = -1$ and
so it is not possible to obtain their action as a tube volume.
Previously, the value $x=-1/3$ has been shown to be an exceptional
case in the context of product spacetime solutions\cite{Maeda-06}.
Also, in the context of the first order theory the value  was found
to be special, since this fine-tuning permitted compactified
solutions with constant torsion on a three-sphere\cite{Canfora-07}.
Here in the context of tube volumes (torsion-free by construction)
we find $\lim_{N\to \infty } x(N) = -1/3$, providing further
evidence that this value is special in some sense. In fact, since a
general 5-manifold may need up to $26$ dimensions in order for an
embedding to exist, we should take $N>>3$ and so $x= -1/3$ to good
approximation.

The Lovelock theory defined by the above action admits two constant
curvature solutions $R^{\mu\nu}_{\ \ \rho\sigma} = \lambda
\delta^{\mu\nu}_{\rho\sigma}$ with Gaussian curvature given by the
roots of a quadratic equation $\lambda = \frac{2(N-1)}{l^2} \left(
-1 \pm \sqrt{1-\frac{N-3}{3(N-1)}}\right)$. So for large $N$ the
characteristic curvature radius of the space-time is given by $l
N^{-1/2}$. Therefore the size of the (5-dimensional) universe is
much smaller then the thickness of the tube. However, in this regime
the tube formula is not valid and so we can not regard the solutions
as meaningful. It can be checked that this appearance of an enormous
effective cosmological constant is a generic feature of for all $D$.
We therefore look for an appropriate term to add to the action,
which may cancel the cosmological constant.

\subsection{The addition of a term proportional to the tube surface area}

If we think of the tube volume as an action functional of the
intrinsic metric, it is interesting to ask what are the extrema of
the action. If one looks for a maximally symmetric solution, with
constant curvature $\lambda$, the Lovelock field equations will give
a polynomial of order $[D/2]$ for $\lambda$. So for $D$ odd one
always has at least one solution. A preliminary investigation
suggests that, for real roots $\lambda$ tends to be large compared
to $1/l^2$ i.e. it describes a geometry where the volume formula is
expected to break down. Also $\lambda$ is always negative.

In order to ensure a maximally symmetric solution with small
curvature, it seems to be necessary to use the tube area formula.
The surface area of the tube is:
 \begin{equation}\label{Weyl_Area_Formula}
 A = \text{Vol}(B_{N-D}) \sum_{n=0}^{[D/2]}
     \frac{(N-D)!!}{(N-D +2n-2)!! (2n)!!} \ { l^{N-D+2n-1}}
     \int_M {\cal L}^{(n)}\, ,
\end{equation}
A more general action would then be $S = \rho V + \sigma A$, with
$\rho, \sigma$ constants. The analogy would be with a drop of fluid,
whose internal energy has an extensive part and also a contribution
from the surface energy.

We shall consider the simple choice
\begin{gather}\label{fundamental}
 S \propto -V + \frac{l}{N-D} A\, .
\end{gather}
This choice allows us to cancel completely the term proportional to
the area of $M$. The resulting action depends only on the curvature
terms,
\begin{gather}\label{not_fundamental}
S \propto \text{Vol}(B_{N-D}) \sum_{n=1}^{[D/2]}
     \frac{(N-D-2)!!}{(N-D +2n)!! (2n-2)!!} \ { l^{N-D+2n}}
     \int_M {\cal L}^{(n)}\, ,
\end{gather}
and therefore Minkowski space will be a solution. More generally,
the absence of the bare cosmological constant ($n=0$) term means
that there will be a branch of the solutions where the curvature is
small compared to $1/l^2$. These solutions will be like solutions of
Einstein's equation with higher order corrections in $l^2$.
Solutions for $M$ belonging to this branch can have tubes that do
not self intersect.

Normalising so that the coefficient of the Einstein-Hilbert term is
unity, the
 coefficients of the Lovelock series are:
\begin{align*}
 \Lambda & = 0\, ,
\\
 \alpha_2 & = \frac{l^{2} }{(N-D+4) 2!! }\, ,
\\
 \alpha_3 & = \frac{ l^{4} }{(N-D+4)(N-D+6) 4!! }\, ,
\end{align*}
etc. Generally
\begin{gather}\label{special_couplings}
 \alpha_n = \frac{  (N-D+2)!! \, l^{2n-2}}{(N-D+2n)!! (2n-2)!! }\, .
\end{gather}

\section{Validity of the tube formula and determinism}

\subsection{Domain of validity of determinism}

To see when determinism breaks down in this theory\footnote{In Ref.
\cite{Choquet-Bruhat-88}\cite{Choquet-Bruhat-88b} determinism is
defined in terms of solving for $\ddot{g}_{ab}$ given initial data
$g_{ab}$ and $\dot{ g}_{ab}$ on a space-like hypersurface. As
discussed in footnote 1, this is not always equivalent to the
Hamiltonian evolution. The former approach will arise naturally when
integrating by finite element approximation. As such it is relevant
to numerical evolution of solutions. The latter approach is more
correct from the point of view of taking limits, for example when we
consider classical solutions as arising from the method of
stationary phase\cite{Gravanis-09}. Here we follow the definition of
Ref. \cite{Choquet-Bruhat-88}\cite{Choquet-Bruhat-88b}, because it
allows us to restore determinism by imposing a simple inequality on
the Riemann tensor. For the Hamiltonian evolution no such simple
condition exists.}, we need to examine the determinant $\det
(1\hspace{-5pt}1 + \mathcal{Y})$. First, in order to simplify the
expression for $\mathcal{Y}$, let us introduce Lovelock tensors,
with two and four free indices, of the spatial components of the
curvature:
\begin{align*}
 (\mathcal {H}^{(p)})^{a}_{\ b} & :=
 -\frac{1}{2^{p+1}}\delta^{a  b_3 \cdots b_{2p}}
 _{b a_1 \cdots a_{2p}} R^{a_1 a_2}_{\quad
 b_1 b_2} \cdots R^{a_{2p-1}a_{2p}}_{\quad b_{2p-1} b_{2p}}\, ,
\\
 (\mathcal {I}^{(p)})^{a\ c}_{\ b \ d} & :=
 -\frac{1}{2^{p+1}}\delta^{a c b_3 \cdots b_{2p}}
 _{b d a_1 \cdots a_{2p}} R^{a_1 a_2}_{\quad
 b_1 b_2} \cdots R^{a_{2p-1}a_{2p}}_{\quad b_{2p-1} b_{2p}}\, .
\end{align*}
Then we obtain the general formula:
\begin{gather}
 \mathcal{Y}_{ab}^{\ \ cd} = -2 \sum_{n=2}^{[D/2]} n \alpha_n \left(
 (\mathcal {I}^{(n-1)})^{\ \ cd}_{ab} - g_{ab} (\mathcal {H}^{(n-1)})^{cd}
 \right)\, .
\end{gather}
Now we shall evaluate this for the choice of coupling coefficients
(\ref{special_couplings}) obtained in the preceeding section. Let us
assume that the embedding space is high dimensional: $N>> D$.
Therefore
\begin{gather*}
 \alpha_n \sim \left(\frac{l^2}{2N}\right)^{n-1} \frac{1}{(n-1)!}\, .
\end{gather*}
\begin{gather}
 \mathcal{Y}_{ab}^{\ \ cd} \sim -2 \sum_{p=1}^{[(D-2)/2]}
 \left(\frac{l^2}{2N}\right)^{p} \frac{1}{p!}  \left(
 (\mathcal {I}^{(p)})^{\ \ cd}_{ab} - g_{ab} (\mathcal {H}^{(p)})^{cd}
 \right)\, .
\end{gather}
The determinant will never vanish if all eigenvalues of
$\mathcal{Y}_I^{\ J}$ are much smaller than unity (in an appropriate
frame, e.g. an orthonormal frame, we may say that all the components
are much less than unity). This will always be the case provided all
$R^{ab}_{\ \ cd} << 2N/l^2$. Determinism will only be in danger of
breaking down once Riemann tensor components become of order
$2N/l^2$.

\subsection{Domain of validity of the tube formula}

As mentioned previously, the tube formula breaks down if the tube
intersects itself in some way. The tube formula is valid provided
every point in the tube has a unique geodesic which connects it with
$M^D$ and intersects $M^D$ normally\footnote{i.e. provided that the
exponential map from the bundle of normal vectors into
$\mathbb{M}^N$ is bijective for normal vectors of length $\leq l$}.
It is easy to check that:

The tube formula in Minkowski space breaks down locally around a
point $x \in M$ if any of the eigenvalues of the extrinsic curvature
matrices $K_{i}(x)$ is greater than or equal to $1/l$ in magnitude.
Furthermore, at least locally, requiring the absolute value of all
the eigenvalues of the $K_i$ to be less than $1/l$ is a sufficient
condition for the validity of the tube formula. In view of the Gauss
relation, this means that the magnitude of components of the Riemann
tensor in an appropriate basis are certainly less than $2N/l^2$. In
fact, since the sectional curvatures will be less than $1/l^2$, the
tube formula is expected to break down when the Riemann tensor
components (in an orthonormal frame say) are of order $1/l^2$.

\subsection{Physical implications}

Although we have only given an order of magnitude estimate, the
result is quite compelling. It gives us strong evidence that the
domain of validity of the tube formula is contained within the
domain of validity of determinism. \emph{If this is so, it means
that in regions where the curvature blows up, the tube formula
breaks down before determinism breaks down.}

When the tube formula breaks down, it is because elements of the
extrinsic geometry interfere with the simple expression of the
volume and area in terms of intrinsic geometry of $M$. If we regard
(\ref{fundamental}) as the fundamental definition of the action,
then formula (\ref{not_fundamental}) is an effective description
only when extrinsic curvatures are small. Once they become large,
there is a phase transition to a regime where the geometrical
degrees of freedom are different.

Therefore, instead of a phase transition to a nondeterministic (and
therefore classically ill-defined) theory, we have a phase
transition to a different sector of the theory where the tube volume
and area are not described entirely in terms of the intrinsic
geometry of $M$, but where extrinsic geometry of the embedding
becomes relevant as a physical degree of freedom.

In all of this we are assuming that in the sector described by
(\ref{not_fundamental}) it is legitimate to vary the action with
respect to the intrinsic metric of $M$, rather than w.r.t. the
embedding itself. This is potentially a rather big weakness, which
we will pick up on again in the concluding section.

\section{Embedding space-times into Minkowski space}

We have treated space-time and its tube as embedded in some
Minkowski space of higher dimension in such a way that the intrinsic
geometry of spacetime coincides with the induced geometry of the
embedding. So far we have just assumed that such isometric
embeddings (of the appropriate level of smoothness) exist. Now it is
necessary to take this question seriously. For Riemannian manifolds,
it is a classic result of J. Nash that any manifold may be smoothly
isometrically embedded into Euclidean space of large enough
dimension. For manifolds of Lorentzian signature, we need to know
what kind of manifolds have such an embedding in $\mathbb{M}^N$.
Fortunately, in recent years a very satisfactory answer to this
question has emerged.

So how does Nash's embedding theorem generalise to Minkowski space?
Clearly, not every space-time admits such an embedding. For example,
if space-time is not time-orientable it can not be
embedded\footnote{It may be possible to embed such a space-time into
a pseudo-Euclidean space $\mathbb{E}^{p , q}$ with $q>1$ time
dimensions. In fact Greene\cite{Greene-70} and
Clarke\cite{Clarke-70} independently showed that any
pseudo-Riemannian manifold can be isometrically embedded into
$\mathbb{E}^{p , q}$ for large enough $p$ and $q$. However, for the
purposes of the tube formula, such embeddings are not acceptable,
due to the problem of defining a tube when there are null geodesics
in the normal space.}. Using straightforward arguments, Penrose
showed that the manifold must admit a spacelike surface separating
space-time into two disconnected regions (past and future), such
that every causal path cuts the surface no more than once and every
timelike curve ending on the surface has bounded proper
time\cite{Penrose-65} (note that this is weaker than global
hyperbolicity- for example take a globally hyperbolic space-time and
remove some points or timelike surfaces. The resulting spacetime
will not be globally hyperbolic but it will still obey the above
condition). A highly non-trivial result - almost the converse of
Penrose's - obtained recently, is the following remarkable
theorem\cite{Mueller-08}:
\begin{theorem}[M\"{u}ller, S\'{a}nchez]\label{iff}
 Any globally hyperbolic space-time manifold $M^D$ admits a global smooth isometric
 embedding into Minkowski space $\mathbb{M}^N$ for large enough $N$.
\end{theorem}
The current upper bound for what is a sufficiently large value of
$N$ is $max\{(D^2 + 5D +2)/2 , (D^2 +3D +12)/2\}$ i.e. one higher
than the corresponding upper bound for Euclidean manifolds. So if we
want to study four-dimensional space-times, we should embed them in
at least 19 dimensions to be sure that an embedding exists. For five
dimensions, we should embed them in 26 dimensions etc.

As mentioned above, global hyperbolicity is not a necessary
condition for the embedding. However, the slightly weaker condition
of causal simplicity is a necessary condition\cite{Mueller-08}.

It is quite wonderful that the existence of the embedding is
guaranteed by only one requirement- and a very welcome requirement
it is too. A globally hyperbolic space-time is the arena for
deterministic physics. This complements rather well the (heuristic)
results of the previous section.

\section{Conclusions and further discussion}

Weyl's formulae for the volume $V$ and surface area $A$ of a tube in
Euclidean space have been shown to generalise straightforwardly to a
tube surrounding a pseudo-Riemannian manifold embedded in Minkowski
space. The resulting formulae correspond to the action of Lovelock
gravity, with a special combination of the coefficients. We have
focussed on just one spacial case, taking a combination $-V +
\frac{l}{N-D} A$ so that the bare cosmological constant term
vanishes from the Lovelock series. In this case evidence was found
that the Lovelock description of the tube volume breaks down
\emph{before} determinism breaks down. Therefore, instead of the
theory itself breaking down, one would have a phase transformation
to a different sector, governed by different geometrical degrees of
freedom. For the future, a more careful study of the curvature
inequalities is needed to verify this, also taking into account
other linear combinations of $V$ and $A$.

In order to obtain the equations of Lovelock gravity, we have not
varied w.r.t. the embedding, but rather the intrinsic geometry of
$M$. This means, for instance that we regard translations of the
tube in Minkowski space as pure gauge. Also, in higher dimensions,
it is possible to have changes in the extrinsic curvature which
preserve the intrinsic metric and curvature. This is known as
\emph{isometric bending}. These are also treated as pure gauge.
However, one is at liberty to question this approach. If the
embedding is regarded as real rather than just metaphorical, the
rigid motions would then be more correctly regarded as zero modes of
the theory. Also the isometric bending would become something like
zero modes. If we follow the string/brane approach in constructing
the variational principle, we should regard the embedding
coordinates $X^A(x^\mu)$ as the degrees of freedom. To derive the
field equations, one follows exactly the same argument as with
standard Regge-Teitelboim geodetic brane
gravity\cite{Regge-77}\cite{Karasik-02}. Noting that $\delta
g_{\mu\nu} =\delta\left( X^A_{,\mu}X^{B}_{ , \nu}\right)\eta_{AB}$
and using the Bach-Lanczos identity $H^{\mu\nu}_{\ \ ;\mu} =0$ we
get:
%\begin{gather}
%\delta \left(\phi^*\left(R^{\mu\nu}\wedge R^{\rho\sigma} \wedge
%(\#e)_{\mu\nu\rho\sigma}\right)\right) = \delta \phi^*
%\left(R^{\mu\nu}\wedge R^{\rho\sigma} \wedge
%(\#e)_{\mu\nu\rho\sigma}\right) + \left(\phi^*\left(R^{\mu\nu}\wedge
%R^{\rho\sigma} \wedge \delta e^\lambda
%(\#e)_{\mu\nu\rho\sigma\lambda}\right)\right)
% +d\left(2\phi^*\left(\delta \omega^{\mu\nu}\wedge R^{\rho\sigma}
%\wedge (\#e)_{\mu\nu\rho\sigma}\right)\right)
%\end{gather}
\begin{equation*}
 \left( \sum_n c_n (H^{(n)})^{\mu\nu} \right) X^A_{;\mu\nu} =0\, .
\end{equation*}
So we see that the solutions of Lovelock gravity would be a subset
of the resulting solutions. However, there are also other solutions
such as the rather trivial $X^A_{;\mu\nu} = 0$. The possible
degeneration of this term multiplying the field equations will
affect any conclusions regarding determinism. Therefore it may be
desirable to avoid varying w.r.t. $X^A(x^\mu)$. More study is
needed.

That there is a formal connection\cite{Labbi-07} between Weyl's tube
formula and Lovelock gravity is, in the authors opinion, of
undoubted interest. It remains to be checked more carefully if this
truly provides a resolution to the problem of indeterminism (or any
other physical problems). In our method there is perhaps some mixing
of philosophies between the realist and the metaphorical
interpretation of the embedding space, which needs to be untangled
in a satisfactory manner. This work is offered as an introduction
and perhaps an invitation to further study of the subject.

\acknowledgements

It is a pleasure to thank J. Zanelli for much useful advice and
inspiration; also thanks to E. Gravanis and R. Troncoso for fruitful
discussions. The Centro de Estudios Cient\'{\i}ficos (CECS) is
funded by the Chilean Government through the Millennium Science
Initiative and the Centers of Excellence Base Financing Program of
Conicyt. CECS is also supported by a group of private companies
which at present includes Antofagasta Minerals, Arauco, Empresas
CMPC, Indura, Naviera Ultragas and Telef\'{o}nica del Sur. This work
was supported in part by Fondecyt grant 1085323.

\end{document}